\documentclass[%
 reprint,
 amsmath,amssymb,
 aps,
]{revtex4-2}

\usepackage{graphicx}% Include figure files
\usepackage[version=3]{mhchem} % Formula subscripts using \ce{}
\usepackage{bm}
\usepackage{amssymb}
\usepackage[colorlinks=true,bookmarks=false,citecolor=blue,urlcolor=blue]{hyperref} 
\usepackage{comment}

\begin{document}

\preprint{APS/123-QED}

\title{Fast multi-channel inverse design through augmented partial factorization}% Force line breaks with \\

\author{Shiyu Li}
\affiliation{Ming Hsieh Department of Electrical and Computer Engineering, University of Southern California, Los Angeles, California 90089, USA}

\author{Ho-Chun Lin}
\affiliation{Ming Hsieh Department of Electrical and Computer Engineering, University of Southern California, Los Angeles, California 90089, USA}

\author{Chia Wei Hsu}
\email{cwhsu@usc.edu}
\affiliation{Ming Hsieh Department of Electrical and Computer Engineering, University of Southern California, Los Angeles, California 90089, USA}

\begin{abstract}
  Computer-automated design and discovery have led to high-performance nanophotonic devices with diverse functionalities.
  However, massively multi-channel systems such as metasurfaces controlling many incident angles and photonic-circuit components coupling many waveguide modes still present a challenge.
  Conventional methods require $M_{\rm in}$ forward simulations and $M_{\rm in}$ adjoint simulations---$2M_{\rm in}$ simulations in total---to compute the objective function and its gradient for a design involving the response to $M_{\rm in}$ input channels.
  %Increasing attention has been devoted to metasurfaces with incident-angle-dependent multifunctionalities, enabling ultra-compact imaging systems superior to conventional optics.
  %Designing metasurfaces with large degrees of freedom remains challenging, especially when multiple inputs are involved.
  %Topology optimization enables efficient design of nanophotonic devices, which requires the objective-function gradient considering all angles of interest.
  %We recently proposed the ``augmented partial factorization'' method to tackle multi-source forward problems by handling numerous inputs simultaneously, but the inverse problem was unsolved.
  By generalizing the adjoint method and the recently proposed augmented partial factorization method, here we show how to obtain both the objective function and its gradient for a massively multi-channel system in a single simulation,
  %to compute such gradient in a single shot, avoiding the slow loop over all inputs as the adjoint method does.
  %More computationally efficient performance can be expected by dividing one such computation into several small ones.
  achieving over-two-orders-of-magnitude speedup and reduced memory usage.
  %Benchmarks show that this approach outperforms the adjoint method by 20 fold in speed and 2 fold in memory for a two-dimensional metasurface with about 4800 inputs.
  We use this method to inverse design a metasurface beam splitter that separates the incident light to the target diffraction orders for all incident angles of interest, a key component of the dot projector for 3D sensing.
  This formalism enables efficient inverse design for a wide range of multi-channel optical systems. 
\end{abstract}

%\setboolean{displaycopyright}{false} % Do not include copyright or licensing information in submission.


\maketitle

\section{Introduction}
%Metasurfaces contain numerous subwavelength meta-atoms that scatter light to different directions and generate desirable wavefronts~\cite{kildishev2013planar,yu2014flat,genevet2017recent,chen2020flat}.
%Conventionally, metasurfaces are designed from a library of unit cells whose transmission coefficients are simulated as a periodic array, following a spatially resolved response ({\it e.g.}, amplitude, phase, polarization) map.
%This computationally cheap approach has given birth to versatile optical components with good performance, such as flat lenses, optical vortex converters and holograms.
%However, this unit-cell-based method becomes less useful as the functionality complexity of the device scales up,
%such as metasurfaces with angle-dependent responses.
Nanoengineered photonic devices can realize versatile and high-performance functionalities in a compact footprint, expanding the limited scope of conventional optical components. %, but require great computational effort of design and simulation.
Computer-automated inverse design~\cite{jensen2011topology,2012_Miller_thesis,molesky2018inverse,elsawy2020numerical,fan2020freeform,li2022empowering,yang2023special} can search a high-dimensional parameter space to discover optimal structures that outperform manual designs or realize new functionalities.
%, such as mode converters, multiplexers, aperiodic meta-structures with angle-dependent responses, {\it etc.}, coming up with a design taking the overall response into consideration remains computationally challenging.
%Topology optimization is a powerful tool for designing nanophotonic devices with superior performance compared to human-designed counterparts
%~\cite{jensen2011topology,molesky2018inverse,elsawy2020numerical,li2022empowering}.
With inverse design, the photonic structure is updated iteratively to optimize an objective function $f$ that encapsulates the desired properties.
Given the many parameters ${\bf p} = \{ p_k\}$ used to parametrize the design, 
efficient optimization typically requires the gradient $\bm{\nabla}_{\bf p} f$ of the objective function with respect to all parameters.
The computational efficiency is a critical consideration since full-wave simulations are necessary to model the complex light-matter interactions at the subwavelength scale, and numerous simulations are needed for the many iterations of the search process.

When the objective function $f$ involves the response to just one input ({\it e.g.}, one incident angle) or one output ({\it e.g.}, one outgoing angle), the adjoint method can compute the complete gradient $\bm{\nabla}_{\bf p} f$ using only one forward and one adjoint simulations~\cite{molesky2018inverse,elsawy2020numerical,fan2020freeform}.
However, when $f$ involves $M_{\rm in}\gg1$ inputs, the adjoint method requires $2M_{\rm in}$ simulations ($M_{\rm in}$ forward, $M_{\rm in}$ adjoint)~\cite{lin2021computational}.
This prohibits the inverse design of large multi-channel systems.
There are many such multi-channel systems including photonic circuits~\cite{bogaerts2020programmable} and aperiodic meta-structures for applications in wide-field-of-view metalenses~\cite{lin2021computational,lin2018topology,li2022thickness,li2023transmission}, beam combiners~\cite{cheng2017optimization,liu2021metasurface}, angle-multiplexed holograms~\cite{kamali2017angle,jang2021independent}, concentrators~\cite{lin2018topology,roques2022toward}, thermal emission control~\cite{rodriguez2011frequency,rodriguez2013fluctuating,yao2022trace},
%A typical example of this multi-input problem is the design of nonlocal metasurfaces~\cite{overvigdiffractive,shastri2022nonlocal} that support angularly diverse responses, with applications in 
image processing~\cite{zhu2017plasmonic,kwon2018nonlocal,guo2018photonic,cordaro2019high,zhou2020flat,zhou2020metasurface,xue2021high}, optical computing~\cite{silva2014performing,zangeneh2021analogue}, space compression~\cite{guo2020squeeze,reshef2021optic,chen2021dielectric}, {\it etc}. %which 
The inverse design of these systems remains challenging.

We recently proposed the ``augmented partial factorization'' (APF) method, which can solve multi-input electromagnetic forward problems in one shot, offering substantial speed-up and memory usage reduction compared to existing methods~\cite{lin2022_APF}.
But the inverse problem was unsolved since the formalism of Ref.~\cite{lin2022_APF} does not yield the gradient.
Here, we generalize APF to enable efficient gradient computation for multi-input problems.
%Both the forward information ({\it i.e.}, the full scattering matrix) and the complete gradient can be computed
We are able to obtain both $f$ and $\bm{\nabla}_{\bf p} f$ in a single or a few computations without a loop over the individual input channels.
%Dividing this large APF computation into several sub-APF computations through a matrix division can further reduce the computing time and memory usage.
For a 1-mm-wide metasurface with 2400 inputs, APF achieves $\sim$150 times speed-up and $\sim$30\% memory usage reduction compared to the conventional adjoint method.
As an example problem, we inverse design a metasurface beam splitter that splits the incident light equally into the $\pm$1 diffraction orders over an incident angular range of 60$^\circ$.
%Such device can be used with a VCSEL array and a microlens array as a dot projector, useful for structured illumination microscopy, 3D endoscopy, and 3D sensing.

\begin{figure*}[ht]
\centering
\includegraphics[width=1\textwidth]{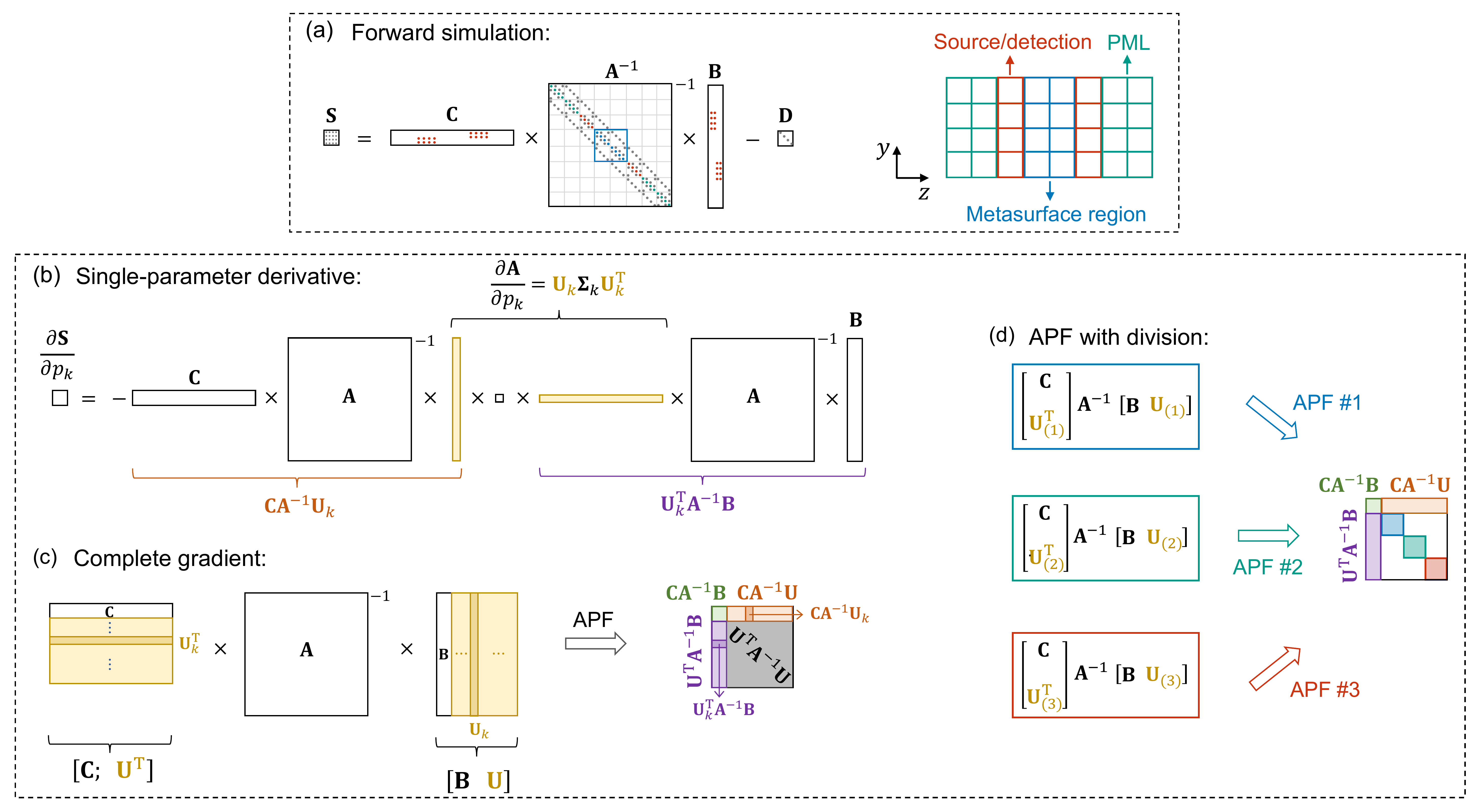}
\caption{Schematics of using augmented partial factorization (APF) for (a) forward simulation and (b--d) gradient computation.
(a) Illustration of $\mathbf{S=CA}^{-1}\mathbf{B-D}$, where the multi-channel response is encapsulated in a generalized scattering matrix $\mathbf{S}$ given in terms of the discretized Maxwell operator matrix $\bf A$, the source profiles $\bf B$, the projection profiles $\bf C$, and the baseline $\bf D$ from the incident field.
Dots represent nonzero elements of these sparse matrices, with colors corresponding to different regions of the simulation domain.
PML: perfectly matched layer.
(b) Illustration of Eqs.~(\ref{eq:dAdp}--\ref{eq:dsdpk_svd}) for a single derivative: after decomposing $\partial \mathbf{A}/\partial p_k$ into $\mathbf{U}_k{\bm \Sigma}_k\mathbf{U}_k^{\rm T}$ and regrouping the terms, $\mathbf{CA}^{-1}\mathbf{U}_k$ and $\mathbf{U}_k^{\rm T}\mathbf{A}^{-1}\mathbf{B}$ can be computed to yield $\partial \mathbf{S}/\partial p_k$.
(c) Illustration of Eq.~(\ref{eq:APF}) for the full gradient: by combining $\mathbf{U}_k$ for all parameters $\{ p_k \}$ and augmenting them to $\mathbf{B}$ and $\mathbf{C}$, a single APF computation can yield the complete gradient with respect to all parameters. 
(d) By dividing $\mathbf{U}$ and performing $N_{\rm sub}$ separate APF computations 
($N_{\rm sub}=3$ in this illustration), %equivalent to separating one large APF computation to $N_{\rm sub}$ small APF computations
we can obtain the same $\mathbf{CA}^{-1}\mathbf{B}$, $\mathbf{CA}^{-1}\mathbf{U}$, and $\mathbf{U}^{\rm T}\mathbf{A}^{-1}\mathbf{B}$ with less computing time and memory by evaluating only $1/N_{\rm sub}$ of the unnecessary matrix $\mathbf{U^{\rm T}A}^{-1}\mathbf{U}$ and by storing smaller matrices $\mathbf{CA}^{-1}\mathbf{U}_{(n)}$ and $\mathbf{U}^{\rm T}_{(n)}\mathbf{A}^{-1}\mathbf{B}$ in memory.
}
\label{fig:Fig1}
\end{figure*}

\section{Multi-input gradient computation using APF}
\label{sec:APF_method}

%\subsection{Multi-input gradient computation using APF}
%\label{subsec:grad_computation}

The novelty of the APF method lies in encapsulating the linear response of the multi-channel system in a generalized scattering matrix $\mathbf{S=CA}^{-1}\mathbf{B-D}$ and then computing the entire $\mathbf{S}$ in a single shot through the partial factorization of an augmented matrix ${\bf{K}} = \left[ {{\bf{A}},{\bf{B}};{\bf{C}},{\bf{D}}} \right]$ that yields its Schur complement $-\mathbf{CA}^{-1}\mathbf{B}$~\cite{lin2022_APF}. Here, the $S_{nm}$ element of the $M_{\rm out} \times M_{\rm in}$ matrix $\mathbf{S}$ is the field amplitude in output channel $n$ given an input in channel $m$ at frequency $\omega$.
Matrix $\mathbf{A}$ is the discretized Maxwell differential operator $- (\omega/c)^2 \varepsilon_{\rm{r}}({\bf{r}}) + \nabla \times \nabla  \times$ of the structure defined by its relative permittivity profile $\varepsilon_{\rm{r}}({\bf{r}})$, the $M_{\rm in}$ columns of matrix $\mathbf{B}$ contain the $M_{\rm in}$ distinct input source profiles, the $M_{\rm out}$ rows of matrix $\mathbf{C}$ contain the $M_{\rm out}$ output projection profiles, and matrix $\mathbf{D}$ subtracts the baseline contribution from the incident field; they are schematically shown in Fig.~\ref{fig:Fig1}a.
If the number of nonzero elements in matrices $\mathbf{B}$, $\mathbf{C}$, and $\mathbf{S}$ are less than the number of nonzero elements in matrix $\mathbf{A}$, the single-shot computing time and memory usage of APF will be independent of how many input and output channels there are~\cite{lin2022_APF}.
In the following, we use APF with finite-difference discretization implemented in our open-source software MESTI~\cite{MESTI}, using the MUMPS package for factorization~\cite{Amestoy2001_SIAM}.
%Figure~\ref{fig:Fig1}a shows a schematic for the forward simulation, yielding the full scattering matrix.

%\clearpage % without this, Fig 1 will get pushed to the end of the paper somehow

Here, we derive a general formulation such that the gradient of any multi-channel objective function can also be computed in a single shot regardless of the number $M_{\rm in}$ of input channels.
We use vector $\mathbf{p} = [p_1, \cdots, p_{N_p} ]$ to denote the $N_p$ real-valued variables parameterizing the photonic design; in the example later, $\{ p_k \}$ will be the edge positions of the ridges of a metasurface.
The objective function $f$ (also called the figure of merit, FoM) that evaluates the performance of the multi-channel device is a function of the generalized scattering matrix $\mathbf{S}$ and the parameters $\mathbf{p}$, namely $f[\mathbf{S}(\mathbf{p}),\mathbf{p}]$.
%For any real-valued FoM with contributions from multiple inputs and outputs ({\it i.e.}, summing over corresponding inputs and outputs in $f[\mathbf{S}(P),P]$) and real-valued optimization variable $p_k$, the 
The gradient $df/dp_k$ we want follows from the chain rule as
\begin{equation}
    \frac{df}{dp_k} = \frac{\partial f}{\partial p_k} + \sum_{n=1}^{M_{\rm out}} \sum_{m=1}^{M_{\rm in}} 2{\rm Re} \left(\frac{\partial f}{\partial S_{nm}}\frac{\partial S_{nm}}{\partial p_k} \right).
    \label{eq:dfdp}
\end{equation}
%where $\mathbf{S}=S_{nm}=S(\theta_{\rm out}^n,\theta_{\rm in}^m)$ represents the scattering matrix, corresponding to the scattered field at angle $\theta_{\rm out}^n$ given an incident plane wave at angle $\theta_{\rm in}^m$;
%$M_{\rm in}$ and $M_{\rm out}$ represent the number of input and output channels ({\it i.e.}, angles) within the angular bandwidth of interest.
Note that $S_{nm}$ is complex-valued, and $\partial f/\partial S_{nm}$ is a Wirtinger derivative.
Both $\partial f/\partial p_k$ and $\partial f/\partial S_{nm}$ can be calculated analytically given the definition of the objective function $f$ for a specific problem.
So, the simulations only need to evaluate the derivative of the scattering matrix, $\partial S_{nm}/\partial p_k$.

The parameters $\{ p_k \}$ modify the scattering matrix $\mathbf{S}$ by modifying the photonic structure $\varepsilon_{\rm{r}}({\bf{r}})$ in the Maxwell operator $\mathbf{A}$. % depends on $p_k$, while the input and output profiles $\mathbf{B}$ and $\mathbf{C}$ are fixed.
Taking the derivative of $\mathbf{S=CA}^{-1}\mathbf{B-D}$ and using the identity $\partial {\bf A}^{-1}/\partial p_k=-{\bf A}^{-1}(\partial {\bf A}/\partial p_k){\bf A}^{-1}$, we get
%\begin{equation}
    $\partial \mathbf{S}/\partial p_k = -\mathbf{CA}^{-1}(\partial \mathbf{A}/\partial p_k)\mathbf{A}^{-1}\mathbf{B}$.
    %\label{eq:dSdp}
%\end{equation}
%as shown in Fig.~\ref{fig:Fig1}b
%since $\partial {\bf A}^{-1}/\partial p_k=-{\bf A}^{-1}(\partial {\bf A}/\partial p_k){\bf A}^{-1}$.
Here we assume that matrices $\mathbf{B}$, $\mathbf{C}$, and $\mathbf{D}$ do not depend on $\{ p_k \}$; additional terms can be added if there is such dependence.
This generalizes the adjoint method to multi-channel systems: the columns of $\mathbf{A}^{-1}\mathbf{B}$ correspond to $M_{\rm in}$ forward problems, and the rows of $\mathbf{CA}^{-1}$ correspond to $M_{\rm out}$ adjoint problems.
To recover the conventional adjoint method, one can substitute $\partial \mathbf{S}/\partial p_k$ into Eq.~(\ref{eq:dfdp}) and sum over the output channels for each input, which converts the $M_{\rm out}$ adjoint problems to $M_{\rm in}$ adjoint problems.
%we do not do so here because that would require the forward problems to be solved prior to the adjoint ones, but here we want to solve the forward and the adjoint problems simultaneously.
%Conventionally, such computation will require $M_{\rm in} + M_{\rm out}$ simulations.

A key observation is that the derivative $\partial \mathbf{A}/\partial p_k$ above is a low-rank matrix since only a few elements of $\mathbf{A}$ depend on the parameter $p_k$.
For example, with 2D transverse magnetic (TM) waves at angular frequency $\omega$, %discretizing with grid size $\Delta x$ yields 
matrix $\mathbf{A}$ is the discretized version of $- {\nabla ^2} - \left( \omega/c \right)^2{\varepsilon _{\rm{r}}}\left({\bf{r}} \right)$, and $\partial \mathbf{A}/\partial p_k$ is zero everywhere except for the few diagonal elements corresponding to pixels where the relative permittivity profile $\varepsilon _{\rm{r}}(\bf r)$ is modified by $p_k$.
Matrix $\partial \mathbf{A}/\partial p_k$ is also symmetric due to reciprocity.
Therefore, we can do a symmetric singular value decomposition
\begin{equation}
    \partial \mathbf{A}/\partial p_k
    %= -(\omega^2/c^2)\Delta x^2\frac{\partial \epsilon_r(\mathbf{r})}{\partial p_k}
    =\mathbf{U}_k\bm{\Sigma}_k\mathbf{U}_k^{\rm T},
    \label{eq:dAdp}
\end{equation}
where $\bm{\Sigma}_k$ is an $r_k$-by-$r_k$ diagonal matrix containing the singular values, $r_k$ is the rank of $\partial \mathbf{A}/\partial p_k$, the $r_k$ columns of $\mathbf{U}_k$ are the left-singular vectors (which are real-valued and equal the right-singular vectors), and $^{\rm T}$ stands for matrix transpose.
%is a low-rank diagonal matrix that is non-zero only at the grid sites where $\epsilon_r(\mathbf{r})$ changes with parameter $p_k$;
%so it can be written in terms of a small number of singular vectors and singular values as $\partial \mathbf{A}/\partial p_k = {\rm diag}([0,\cdots,\sigma_n^k,\cdots,0])=\sum_{n=1}^{N_{\sigma}^k} \sigma_n^k u_n^k (v_n^k)^{\dagger}$, with ${\bf U}^k = {\bf V}^k=[u_1^k,\cdots,u_{N_{\sigma}^k}^k]=[v_1^k,\cdots,v_{N_{\sigma}^k}^k]$;
%$\sigma_n^k$ equals $\partial \mathbf{A}/\partial p_k$ at the corresponding pixel specified by column vectors $u_n^k=v_n^k=[0;\cdots;0;1;0;\cdots;0]$, corresponding to the $n$-th pixel changed by $p_k$;
These singular vectors are zero everywhere except at the pixels where $\varepsilon _{\rm{r}}(\bf r)$ is modified by $p_k$.
%$N_{\sigma}^k$ is the number of non-zero singular values, equaling the number of pixels changed by $p_k$.
%Doing singular value decomposition (SVD) this way yields sparse matrices ${\bf U}^k$ and ${\bf V}^k$, whose sparsity reduces the computing time and memory usage of APF.

We then obtain the derivative of the scattering matrix with respect to the $k$-th parameter
\begin{equation}
    \partial \mathbf{S}/\partial p_k=-(\mathbf{CA}^{-1}\mathbf{U}_k) \bm{\Sigma}_k (\mathbf{U}_k^{\rm T}\mathbf{A}^{-1}\mathbf{B}),
    \label{eq:dsdpk_svd}
\end{equation}
as shown in Fig.~\ref{fig:Fig1}b.
%Note that matrices $\mathbf{CA}^{-1}\mathbf{U}^k$ and $\mathbf{V}^{k\dagger}\mathbf{A}^{-1}\mathbf{B}$ can each be computed in a single shot using APF without looping over the $M_{\rm in}$ input channels, but this only yields the gradient for one parameter, $p_k$.
To obtain the complete gradient with respect to all $N_p$ parameters $\{ p_k \}$, we combine the singular-vector matrices as $\mathbf{U} \equiv [\mathbf{U}_1,\cdots,\mathbf{U}_{N_p}]$, % and $\mathbf{V}=[\mathbf{V}^1,\cdots,\mathbf{V}^{N_p}]$.
which has $M_p \equiv \sum_{k=1}^{N_p} r_k$ columns.
Computing % $\mathbf{CA}^{-1}\mathbf{B}$, 
$\mathbf{CA}^{-1}\mathbf{U}=[\mathbf{CA}^{-1}\mathbf{U}_1,\cdots,\mathbf{CA}^{-1}\mathbf{U}_{N_p}]$ and $\mathbf{U}^{\rm T}\mathbf{A}^{-1}\mathbf{B}=[\mathbf{U}_1^{\rm T}\mathbf{A}^{-1}\mathbf{B};\cdots;\mathbf{U}_{N_p}^{\rm T}\mathbf{A}^{-1}\mathbf{B}]$ with APF would yield the complete gradient through Eqs.~(\ref{eq:dfdp}--\ref{eq:dsdpk_svd}) with just two APF computations.
As shown in Fig.~\ref{fig:Fig1}c, we can further reduce from two APF computations to one by building new matrices $\mathbf{\tilde{B}=[B, U]}$ and $\mathbf{\tilde{C}=[C; U^{\rm T}]}$ and using APF to compute %$\mathbf{\tilde{C}A}^{-1}\mathbf{\tilde{B}}$, which has components
\begin{equation}     \mathbf{\tilde{S}}=\mathbf{\tilde{C}A}^{-1}\mathbf{\tilde{B}} = 
    \begin{aligned}
    \left[ \begin{matrix} \mathbf{C} \\ \mathbf{U^{\rm T}} \end{matrix} \right]\mathbf{A}^{-1} \left[ \begin{matrix} \mathbf{B} & \mathbf{U} \end{matrix}\right]
    = \left[ \begin{matrix} \mathbf{CA}^{-1}\mathbf{B} & \mathbf{CA}^{-1}\mathbf{U} \\ \mathbf{U^{\rm T}A}^{-1}\mathbf{B} & \mathbf{U^{\rm T}A}^{-1}\mathbf{U} \end{matrix} \right]
\end{aligned}
\label{eq:APF}
\end{equation}
Here, the matrix $\mathbf{\tilde{K}} = \left[ {{\bf{A}},\mathbf{\tilde{B}};\mathbf{\tilde{C}},{\bf{0}}} \right]$
is augmented with not only the original $M_{\rm in}$ input and $M_{\rm out}$ output channel profiles $\mathbf{B}$ and $\mathbf{C}$ but also $M_p$ additional inputs/outputs being the singular vectors $\mathbf{U}$ and $\mathbf{U}^{\rm T}$ from the design parameters $\{p_k\}$.
This way, a single-shot APF computation solves all of the $M_{\rm in}$ forward simulations (yielding the scattering matrix ${\bf S}$ from $\mathbf{CA}^{-1}\mathbf{B}$) and also obtains the complete gradient ({\it i.e.}, $\partial \mathbf{S}/\partial p_k$ and $df/dp_k$ for all $p_k$, from $\mathbf{CA}^{-1}\mathbf{U}$ and $\mathbf{U^{\rm T}A}^{-1}\mathbf{B}$) at the same time.
%In most problems, only a subset of the scattering matrix is concerned, such as the transmission matrix $\bf T$ for a metasurface, and the APF-based gradient computation can be easily generalized to only include desired input and output channels.

%\subsection{Speed-up through matrix division}
%\label{subsec:matrix_division}

While a single APF computation suffices, doing so is not necessarily the most computationally efficient.
When the number of elements in matrix $\mathbf{\tilde{S}}$, $\mathrm{numel}(\mathbf{\tilde{S}}) = (M_{\rm in}+M_p)(M_{\rm out}+M_p)$, is less than the number of nonzero elements in the Maxwell operator matrix $\mathbf{A}$, the APF computing time and memory usage are independent of how many inputs and outputs there are (including the singular vectors $\mathbf{U}$), and we can include as many inputs/outputs as we want in a single APF computation.
But when $\mathrm{numel}(\mathbf{\tilde{S}}) > \mathrm{nnz}(\mathbf{A})$, the APF computing time and memory usage will grow linearly with $\mathrm{numel}(\mathbf{\tilde{S}})$~\cite{lin2022_APF}; this may be the case here since topology optimization often includes a large number of parameters.
We can mitigate this increase by avoiding the computation of
%The most time-and-memory-consuming stage of this single APF computation is the partial factorization of the augmented matrix $\mathbf{K}$.
%Since only matrices $\mathbf{CA}^{-1}\mathbf{B}$, $\mathbf{CA}^{-1}\mathbf{U}$ and $\mathbf{V^{\dagger}A}^{-1}\mathbf{B}$ in Eq.~(\ref{eq:APF}) are needed for the evaluation of $f$ and $\bm{\nabla}_P f$, computing the much larger matrix 
the $M_p$-by-$M_p$ matrix $\mathbf{U^{\rm T}A}^{-1}\mathbf{U}$ in Eq.~(\ref{eq:APF}) [gray-shaded area in Fig.~\ref{fig:Fig1}c], which is not needed for either the scattering matrix or the gradient.
%We can reduce computing time and memory usage by avoiding the computation of $\mathbf{U^{\rm T}A}^{-1}\mathbf{U}$.
As illustrated in Fig.~\ref{fig:Fig1}d, we do so by dividing the singular-vector matrix $\mathbf{U}$ into $N_{\rm sub}$ submatrices, $\mathbf{U}=[\mathbf{U}_{(1)},\cdots,\mathbf{U}_{(N_{\rm sub})}]$ and separating the single APF computation into $N_{\rm sub}$ sub-APF computations, each operating on smaller matrices $\mathbf{\tilde{B}}_{(n)}=[\mathbf{B}, \mathbf{U}_{(n)}]$ and $\mathbf{\tilde{C}}_{(n)}=[\mathbf{C}; \mathbf{U}_{(n)}^{\rm T}]$.
This way, only $1/N_{\rm sub}$ of the unnecessary matrix $\mathbf{U^{\rm T}A}^{-1}\mathbf{U}$ [areas shaded in red, green, and blue in Fig.~\ref{fig:Fig1}d] is computed, reducing computing time and memory usage.
To minimize memory, one can choose $N_{\rm sub}$ to reduce $\mathrm{numel}(\mathbf{\tilde{S}}) \approx [M_{\rm in}+(M_p/N_{\rm sub})]^2$ of each sub-APF computation to the order of magnitude of $\mathrm{nnz}(\mathbf{A})$.
One may merge the output of the $N_{\rm sub}$ computations to obtain $\mathbf{CA}^{-1}\mathbf{U}=[\mathbf{CA}^{-1}\mathbf{U}_{(1)},\cdots,\mathbf{CA}^{-1}\mathbf{U}_{(N_{\rm sub})}]$ and similarly $\mathbf{U}^{\rm T}\mathbf{A}^{-1}\mathbf{B}=[\mathbf{U}^{\rm T}_{(1)}\mathbf{A}^{-1}\mathbf{B};\cdots;\mathbf{U}^{\rm T}_{(N_{\rm sub})}\mathbf{A}^{-1}\mathbf{B}]$, but there is no such need:
we can directly apply Eq.~(\ref{eq:dsdpk_svd}) onto $\mathbf{CA}^{-1}\mathbf{U}_{(n)}$ and $\mathbf{U}^{\rm T}_{(n)}\mathbf{A}^{-1}\mathbf{B}$ without merging them.
By storing $\mathbf{CA}^{-1}\mathbf{U}_{(n)}$ instead of the entire $\mathbf{CA}^{-1}\mathbf{U}$, we can further reduce memory usage.
%This way, only $1/N_{\rm sub}$ of the unnecessary matrix $\mathbf{U^{\rm T}A}^{-1}\mathbf{U}$ [areas shaded in red, green, and blue in Fig.~\ref{fig:Fig1}d] is computed, reducing computing time and memory usage.
%As discussed in Sec.~1 of the Supporting Information, 
%the number of subsets $N_{\rm sub}$ that minimizes the computational cost depends on the specific problem, namely the size of matrix $\mathbf{\tilde{S}}$.
%When the number of elements in matrix $\mathbf{\tilde{S}}$ is larger than the number of non-zero elements in matrix $\bf A$, {\it i.e.}, numel($\mathbf{\tilde{S}})=M_{\rm in}M_{\rm out}>$ nnz($\bf A$), computing and storing it using APF take $\mathcal{O}(M_{\rm in}M_{\rm out})$ time and memory.
%If matrix division reduces the size of $\mathbf{\tilde{S}}$ to numel($\mathbf{\tilde{S}})\approx M_{\rm in}M_{\rm out}/N_{\rm sub}^2<$ nnz($\bf A$), where the computational cost is dominated by the metasurface size rather than $M_{\rm in}$ and $M_{\rm out}$,
%it is possible to significantly reduce the timing and memory usage for problems with extremely large $M_{\rm in}$ and $M_{\rm out}$ ({\it e.g.}, gradient computation).
%For the metasurface design considered in this work, $N_{\rm sub}$ resulting in nnz($\bf \Tilde{S}$)/nnz($\bf A$)$\approx$ 5 is a reasonable estimate, where nnz($\cdot$) takes the number of non-zero elements.

This formalism for computing the gradient of a multi-channel objective function is very general. It applies to any dimension, polarization, discretization scheme, any type of input source profiles and output projection profiles, with any objective function and any set of design variables.

\begin{figure*}[t]
\centering
\includegraphics[width=1\textwidth]{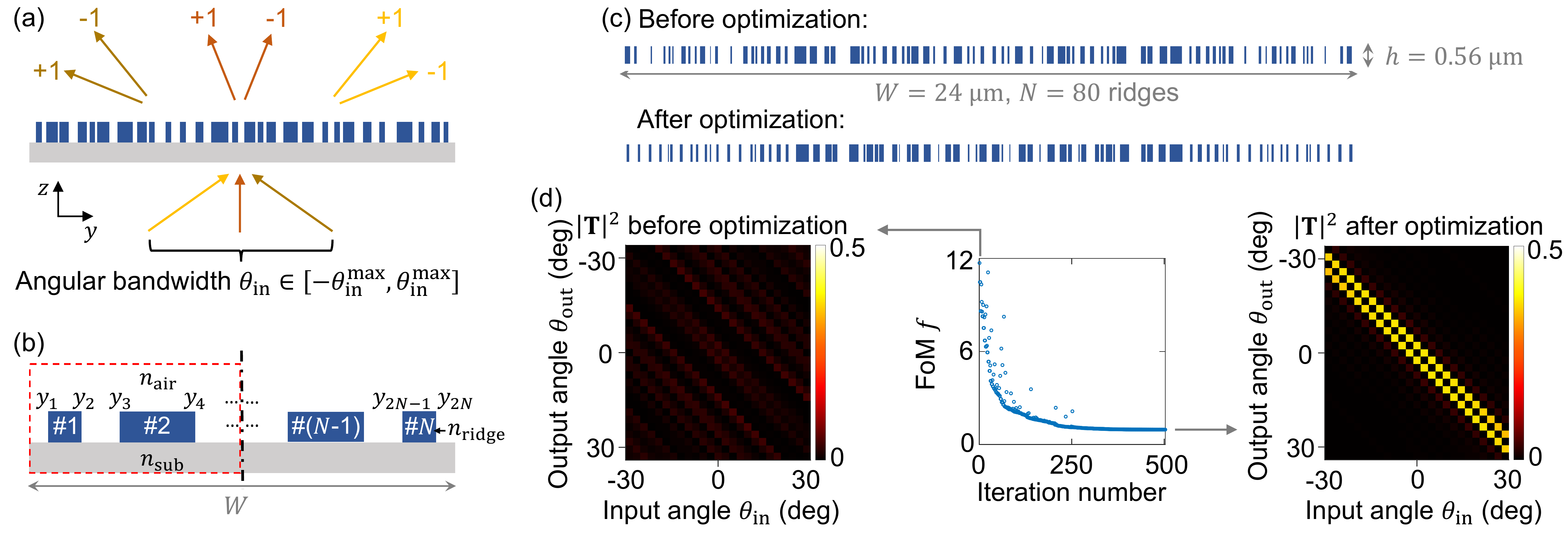}
\caption{Inverse design of a broad-angle metasurface beam splitter.
    (a) The desired broad-angle beam-splitting response.
    (b) Parameter definition and mirror symmetry of the structure.
    (c) Metasurfaces before and after optimization.
    (d) Evolution of the FoM of Eq.~(\ref{eq:FOM_deflector}) during optimization and the squared amplitude of the transmission matrices before and after optimization.}
\label{fig:Fig2}
\end{figure*}

As a concrete example, we consider full-wave modeling of the TM fields of a 1200-ridge aperiodic metasurface in 2D with mirror symmetry regarding its central plane ($W = 1200\lambda$ wide, $L = 0.6\lambda$ thick, discretized with grid size $\Delta x = \lambda/40$ where $\lambda$ is the wavelength), computing its full transmission matrix with $M_{\rm in}=M_{\rm out}=2W/\lambda = 2400$ plane-wave channels on each side and the gradient of the transmission matrix with respect to $N_p = 1200$ parameters being the edge positions of the ridges within half of the metasurface.
Here, $\mathrm{nnz}(\mathbf{A}) \approx 1.6\times 10^7$, and $M_p = 57600$.
%As detailed in Sec.~1 of the Supporting Information, for a 4000-ridge metasurface with 4798 inputs and outputs,  
Compared to the conventional adjoint method, APF with $N_{\rm sub}=15$ reduces the gradient evaluation time from 1354 minutes to 9.5 minutes and the memory usage from 10 GiB to 7 GiB.
%Further increasing $N_{\rm sub}$ may lead to a longer computation, but the memory usage can reduce by up to 50\%, which is extremely useful when memory becomes a concern.
%For this metasurface, the full scattering matrix is computed 10 times; the average computing time and the maximal recorded memory usage after subtracting the 0.95 GiB memory used by MATLAB R2022a among them are used.
Here, the conventional adjoint simulations are also performed with software MESTI~\cite{MESTI}, which is already optimized by using the efficient MUMPS package~\cite{Amestoy2001_SIAM}, utilizing the symmetry of the Maxwell operator matrix ${\bf A}$ and the sparsity of the input profiles.
%and reusing the LU factors for all of the forward and adjoint computations.
All computations run on a single core on an Intel Xeon 2640v4 node.
Details are provided in Sec.~1 of the Supplement~1, Table S1 shows the breakdown of the computing times, and Fig.~S1 plots the dependence on $N_{\rm sub}$.
We have made our gradient computation code open-source~\cite{APF_inv_github}, including both the APF version and the conventional adjoint version.

\section{Inverse design of a broad-angle metasurface beam splitter}
\label{sec:metasurface_splitter}

As an example, here we inverse design a 2D metasurface beam splitter for TM polarization, composed of ridges with different widths, shown in Fig.~\ref{fig:Fig2}a. We want the metasurface to split the incident light equally into the $\pm 1$ diffraction orders for any incident angle $\theta_{\rm in}$ within a $2\theta_{\rm in}^{\rm max}$ angular range.
Such a broad-angle beam splitter can be used with a vertical-cavity surface-emitting laser (VCSEL) array and a microlens array as a dot projector to generate structured illumination useful for structured illumination microscopy~\cite{gustafsson2000surpassing,gustafsson2005nonlinear}, 3D endoscopy~\cite{furukawa20163d,furukawa2016shape}, and 3D sensing ({\it e.g.,} facial recognition~\cite{kwon2021stereoscopic}, motion detection~\cite{dutta2012evaluation}).
The microlens array collimates the output from the VCSEL array, and the beam splitter increases the number of dots.
VCSEL arrays are widely used for dot projectors due to their uniform intensity pattern, high power density, low cost, and simple packaging.
However, existing beam splitters based on Dammann gratings~\cite{li2015all,yang2017simple} or metasurfaces~\cite{li2018full,song2018selective,ni2020metasurface} only operate on normal incident light and not the %rely on the deflection of a normally incident light to several diffraction orders to produce the desired dot patterns, making it difficult to work under 
oblique incidence from the off-axis VCSEL units.

We let the metasurface be periodic with the width of one period being $W=24$ \textmu m, which couples transverse angular momenta $k_y$ differing by integer multiples of $2\pi/W$.
Within the width $W$, we place $N=80$ amorphous-silicon ridges ($n_{\rm ridge}=3.70$) with height $h=0.56$ \textmu m and varying widths and positions, sitting on a silica ($n_{\rm sub}=1.45$) substrate and surrounded by air ($n_{\rm air}=1$), as shown in Fig.~\ref{fig:Fig2}b,c.
The ridge height ensures a sufficient 2$\pi$ range of phase shifts when varying % with high transmission over the angular range $-30^{\circ}\sim 30^{\circ}$ is achieved by changing 
the ridge width (Supplementary Sec.~2 and Fig.~S2).
Since the desired response is symmetric, we let the structure be mirror symmetric with respect to a central plane at $y=0$ (black dot-dashed line). %, as shown in Fig.~\ref{fig:Fig3}b.
The operating wavelength is $\lambda=940$ nm, typical for VCSELs. The angular range is $2\theta_{\rm in}^{\rm max} = 60^{\circ}$, typical for dot projectors.
%Within the width of $W=24\lambda$ in one period, there are $N=80$ amorphous-silicon ridges ($n_{\rm ridge}=3.70$) with height $h=0.56$ \textmu m and varying widths and positions, sitting on a silica ($n_{\rm sub}=1.45$) substrate and surrounded by air ($n_{\rm air}=1$), as shown in Fig.~\ref{fig:Fig3}b,c.
%The ridge height ensures a sufficient 2$\pi$ range of phase shifts when varying % with high transmission over the angular range $-30^{\circ}\sim 30^{\circ}$ is achieved by changing 
%the ridge width (Supporting Information Sec.~2 and Fig.~S3).

To inverse design this broad-angle beam splitter, we minimize the following FoM
\begin{equation}
    f[{\bf T}({\bf p}),{\bf p}] = \sum_{n=1}^{M_{\rm out}} \sum_{m=1}^{M_{\rm in}}\left||T_{nm}({\bf p})|^2-T_{nm,{\rm target}}^2\right|^2,
    \label{eq:FOM_deflector}
\end{equation}
where $T_{nm}$ is the transmission coefficient from the $m$-th incident angle to the $n$-th outgoing angle.
Here, ${\bf p} = \{ p_k\} = \{ y_1,\cdots,y_N\}$ parameterize the two edge positions of the $N/2$ ridges within half a period of the metasurface [encircled by the red box in Fig.~\ref{fig:Fig2}b], with $\{ y_{N+1},\cdots,y_{2N}\}=-\{ y_N,\cdots,y_1\}$ based on symmetry.
The target transmission is $T_{nm,{\rm target}}^2=0.5$ at the $\pm 1$ diffraction orders and $T_{nm,{\rm target}}^2=0$ otherwise.
Equation~(\ref{eq:FOM_deflector}) sums over all incident angles within the angular range of interest [{\it i.e.}, $|k_y^m|<(2\pi/\lambda)\sin{\theta_{\rm in}^{\rm max}}$ with $2\pi/W$ spacing] and all outgoing angles.
With $W=24$ \textmu m, we have $M_{\rm in}=25$ input channels within the $60^{\circ}$ angular range and $M_{\rm out}=51$ output channels.
For this specific FoM, $\partial f/\partial p_k=0$ and $\partial f/\partial T_{nm}=2(|T_{nm}|^2-T_{nm,{\rm target}}^2)T_{nm}^*$ with $^*$ standing for complex conjugation.
%As discussed in Sec.~\ref{sec:APF_method}, dividing a large-sized APF computation into $N_{\rm sub}$ small-sized APF computations can potentially reduce the factorization time and speed up the gradient calculation.
Here, $\mathrm{nnz}(\mathbf{A}) \approx 3.3\times 10^5$, and $M_p = 3840$.
Using the APF method above, each objective-plus-gradient evaluation with $N_{\rm sub}=3$ takes 1.6 seconds while using 0.2 GiB of memory when running on one core.
%A breakdown of the computing time is summarized in Tab.~S2 of the Supporting Information, and 
%The dependence on $N_{\rm sub}$ is shown in Fig.~S2 of the Supporting Information.
%It is about 1.1 times faster compared to the adjoint method.
%Thus, we choose $N_{\rm sub}=3$ to compute the gradient and perform gradient-based optimizations for the beam splitter.
To validate that there is no mistake in our derivation and implementation, we show in Fig.~S3 that the gradient computed with APF agrees with a brute-force finite-difference evaluation of the FoM in Eq.~(\ref{eq:FOM_deflector}).

%Metasurfaces that deflect light as expected have small FoM.
To minimize the FoM, we use gradient-based algorithms to update optimization variables $\bf p$ along the opposite direction of the gradient $\bm{\nabla}_{\bf p} f$.
The optimizations stop when the change of $f$ is less than $f_{\rm tol}^{\rm abs}=10^{-4}$.
After comparing four different algorithms (Supplementary Sec.~4 and Fig.~S4), we choose the sequential least-squares quadratic programming (SLSQP) algorithm~\cite{kraft1994algorithm} implemented in the open-source package NLopt~\cite{Nlopt} since it typically converges the fastest and often to a lower local minimum.
During the optimization, the separation between neighboring edges (both the ridge width and the spacing between ridges) is constrained to be at least 40 nm to ensure fabrication feasibility.

\begin{figure}[tbp]
\centering
\includegraphics[width=0.27\textwidth]{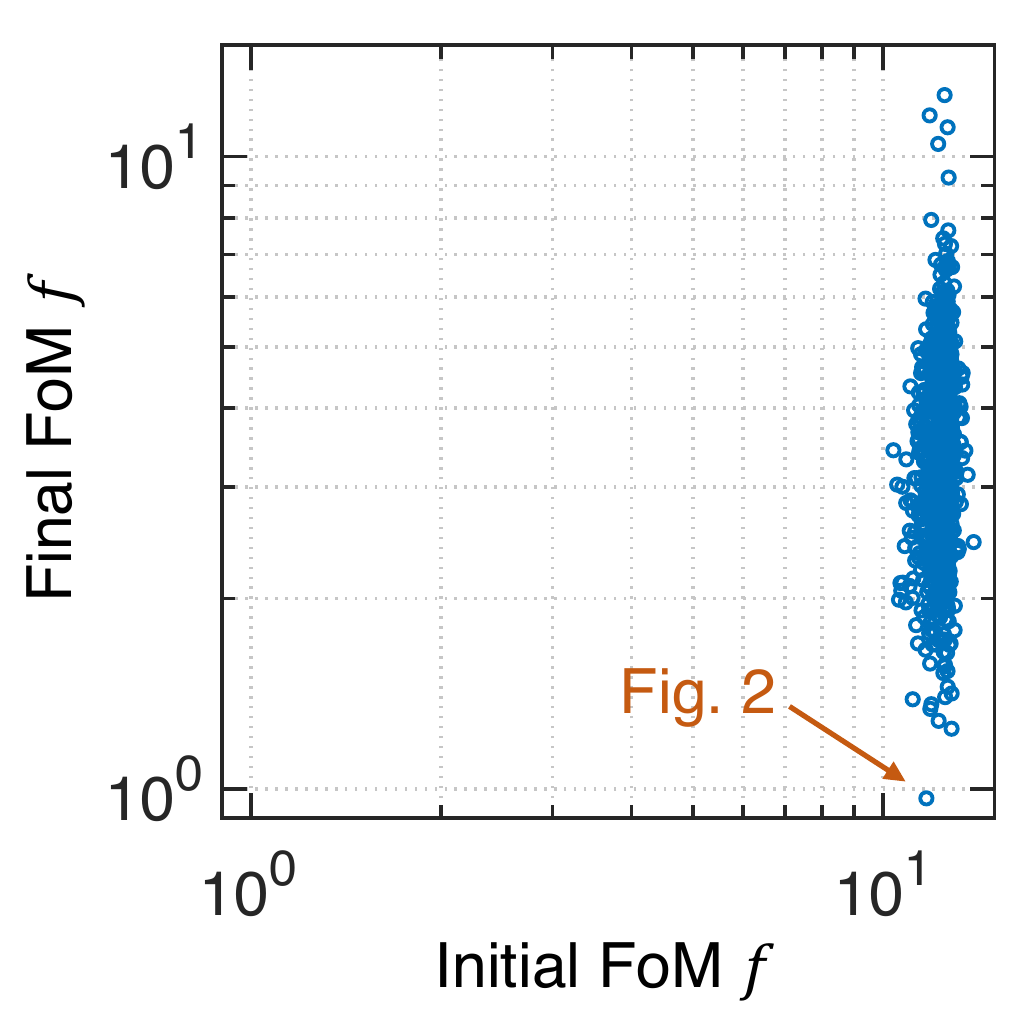}
\caption{The final FoM and the initial FoM for 1000 randomly generated initial configurations.}
\label{fig:Fig3}
\end{figure}

%we then run SLSQP optimizations to 1000 
We find that randomly sampled configurations of the parameter $\bf p$ have a poor performance with the FoM narrowly distributed between 10 and 20, but SLSQP optimization using those configurations as the initial guess leads to a wide distribution of the optimized FoM (Fig.~\ref{fig:Fig3}).
Since this inverse-design problem is non-convex, there is a sensitive dependence on the initial guess. %FoM can not promise a small final FoM since no relation exists between them (see Fig.~S5 of the Supporting Information).
To find a good final design, we run SLSQP optimizations with 1000 different initial guesses.
%These optimizations stop after convergence when the change of $f$ is less than $f_{\rm tol}^{\rm abs}=10^{-4}$ for an update.

Figure~\ref{fig:Fig2}c,d shows the initial configuration, the final configuration, their corresponding transmission matrices, and the evolution of the FoM for the best case among these optimizations.
%As the optimization goes on, the FoM reduces significantly and converges after 503 iterations.
The optimized metasurface exhibits uniform and near-perfect beam splitting for all incident angles within the $60^{\circ}$ angular range of interest.
Here, $|T_{nm}|^2$ averages to 0.4 at the $\pm 1$ diffraction orders and to 0.003 away from these angles.
Supplementary Video 1 shows how the configuration and the transmission matrix evolve, and Table S2 lists the parameters of the final configuration. % in each iteration, from which an obvious splitting is observed after about 50 iterations.
%Without inverse design, it is not obvious whether such a broad-angle beam-splitting functionality can be achieved.
%But our design shows that it is possible.

In Sec.~6 and Figs.~S6--S7 of the Supplement~1, we consider optimizations where we do not impose a mirror symmetry in the structure.
While such a setup is more general, it has a larger-than-necessary design space and yields slightly less optimal results.

\begin{comment}
We also consider general metasurfaces composed of $N$ ridges in each period without mirror symmetry.
Then $P = \{ p_k\} = \{ y_1,\cdots,y_{2N}\}$ parameterizes the two edge positions of all $N$ ridges of the metasurface.
The same stop criterion and constraints of the variables are applied.
The best results obtained from 1000 randomly generated initial guesses are shown in Fig.~S7 of the Supporting Information.
The optimized metasurface still performs much better than the initial guess, but with $|T_{nm}|^2$ averaged to 0.3 at the $\pm 1$ diffraction orders over the $60^{\circ}$ angular range, which is not so good as the optimized symmetric metasurface in Fig.~\ref{fig:Fig3}.

Figure~\ref{fig:Fig4} and Supplementary Fig.~S8 summarize the final FoM and optimization timing from 1000 different initial designs, for symmetric and general metasurfaces respectively.
The average FoM and optimiztion time suggest that symmetric metasurfaces tend to converge to smaller FoM while requiring fewer iterations compared to general metasurfaces, which makes intuitive sense since the desired response is symmetric.
The best performance comes from the symmetric metasurface shown in Fig.~\ref{fig:Fig3}c,d, with a minimized FoM of 0.97 after a 16-min-long optimization

\begin{figure}[t]
\centering
\includegraphics[width=0.35\textwidth]{fig4.pdf}
\caption{The final FoM and optimization time starting from 1000 different symmetric metasurfaces, with dashed lines representing the averaged FoM (red) and optimization time (yellow).}
\label{fig:Fig4}
\end{figure}
\end{comment}

\section{Outlook}

%Although we use 2D TM modes as an example to demonstrate how to use APF to efficiently compute the gradient, the formalism introduced in Sec.~\ref{sec:APF_method} also works for transver-electric (TE) polarization.
%The only difference is $\partial {\bf A}/\partial p_k$, which is diagonal for TM modes and non-diagonal for TE modes.
%Each case we can do a singular value decomposition [Eq.~(\ref{eq:dAdp})] and reorganize terms to the form of ${\bf CA}^{-1}{\bf B}$ [Eq.~(\ref{eq:dsdpk_svd})] to support APF computations.

Computer-automated design and discovery unlock numerous possibilities, and the formalism proposed here can be the enabling element for inverse design on a wide range of multi-channel optical systems mentioned in the introduction.
%This efficient objective-plus-gradient evaluation is useful for any 
%multi-input inverse-design problem described by a scattering-matrix-related, real-valued FoM, such as
%These include beam combiners~\cite{cheng2017optimization,liu2021metasurface}, angle-multiplexed holograms~\cite{kamali2017angle,jang2021independent}, wide-field-of-view metalenses~\cite{lin2021computational,lin2018topology,li2022thickness,li2023transmission}, concentrators~\cite{lin2018topology,roques2022toward}, multi-channel photonic circuits~\cite{bogaerts2020programmable}, and thermal emission~\cite{rodriguez2011frequency,rodriguez2013fluctuating,yao2022trace}.
%It can also be easily adopted to solve for the gradient of other objective functions beyond photonics that can be written as ${\bf CA}^{-1}{\bf B}$.
%We can benefit even more from this method in both timing and memory for large-area systems.
Instead of the $N_{\rm sub}$ matrix division employed here, future work may also explore computing the Schur complement of a rectangular augmented matrix to avoid computing the unnecessary matrix $\mathbf{U^{\rm T}A}^{-1}\mathbf{U}$.
Advances in the computation method, together with open-source codes, can deliver the next generation of photonic devices.

%This formalism can be generalized to three dimensional (3D) systems, presumably saving more time and memory compared to the conventional adjoint method.
%The APF codes in 3D are under development, which may enable the computationally efficient simulation and inverse design of 3D devices with angle-dependent responses.

\section*{Acknowledgments}
The authors thank H. Tahara, S. Komori, A. Akiba, and M. Torfeh for useful discussions.
This work is supported by the National Science Foundation CAREER award (ECCS2146021) and the Sony Research Award Program. Computing resources are provided by the Center for Advanced Research Computing at the University of Southern California.
\bf{Disclosures:} \rm The authors declare no conflicts of interest.
\bf{Data Availability Statement:} \rm All data needed to evaluate the conclusions in this study are presented in the paper and supplemental document. The code is available on GitHub~\cite{APF_inv_github}.
\bf{Supplemental document:} \rm
See Supplement 1 and Video 1 for supporting content.

\begin{comment}
    
\begin{suppinfo}

Benchmarks for metasurfaces with different sizes;
Phase and amplitude maps of ridges for different widths and incident angles;
Comparison of various gradient-based algorithms;
Comparison of final FoM and initial FoM;
Broad-angle metasurface beam splitter designed with general metasurfaces without mirror symmetry (PDF)

Evolution of the metasurface, its transmission matrix and the FoM as the optimization goes on (MPEG)

\end{suppinfo}
\end{comment}

\bibliography{bib_APF_inverse_design}

\end{document}

%\onehalfspacing
\maketitle

\tableofcontents

\section{Benchmarks on gradient computation}
\label{sec:benchmark}

As an example, here we compare the computing time and memory usage of $f$ only, %[obtained from computing only ${\bf CA}^{-1}{\bf B}$],
and $f$-plus-$\bm{\nabla}_\mathbf{p} f$ computation
%[obtained from computing ${\bf CA}^{-1}{\bf B}$, ${\bf CA}^{-1}{\bf U}$, ${\bf U^{\rm T}A}^{-1}{\bf B}$ in Eq.~(4) of the main text]
for a 1-mm-wide metasurface, using the augmented partial factorization (APF)~\cite{lin2022_APF} and conventional adjoint method.
The metasurface (1200$\lambda$ wide) contains 1200 amorphous-silicon ridges with height 0.6$\lambda$ sitting on a silica substrate; it has mirror symmetry with respect to the central plane such that the $N_p=1200$ edge positions of ridges on the left-half side of the metasurface can be changed independently.
We evaluate the full-wave responses with  $M_{\rm in}=M_{\rm out}= 2W/\lambda = 2400$ inputs and outputs using the open-source software MESTI~\cite{MESTI}.

\begin{table}[htbp]
\newcommand{\tabincell}[2]{\begin{tabular}{@{}#1@{}}#2\end{tabular}}
\centering
\setlength\extrarowheight{-3pt}
\caption{\bf Computing time and memory usage for a 1-mm-wide metasurface}
\begin{tabular}{lcccc}
\toprule
    \multicolumn{5}{c}{\tabincell{c}{$N=1200$ ridges \\ 2399 input angles, 2399 output angles}} \\
\midrule
    & \multicolumn{2}{c}{$f$ only} & \multicolumn{2}{c}{$f$ \& $\bm{\nabla}_\mathbf{p} f$} \\
\cmidrule(rr){2-3} \cmidrule(rr){4-5}
   Method & APF & Conventional & \tabincell{c}{APF \\ $N_{\rm sub}=15$} & \tabincell{c}{Conventional \\ adjoint} \\
\midrule
    \multicolumn{5}{c}{Computing time (minute)} \\
\midrule
  Build & 0.03 & 58 & 0.4 & 113 \\
  Analyze & 0.05 & 127 & 0.8 & 249 \\
  Factorize & 0.24 & 395 & 5.7 & 788 \\
  Solve & 0 & 95 & 0 & 192 \\
  Decompress & 0.01 & 0 & 0.3 & 0 \\
  Post-process & 0.0 & 0.0 & 2.3 & 12 \\
  Total & 0.33 & 675 & 9.5 & 1354 \\
\midrule
    \multicolumn{5}{c}{Memory usage (GiB)} \\
\midrule
    & 4 & 10 & 7 & 10 \\
\bottomrule
\end{tabular}
  \label{tab:benchmark_1200}
\end{table}

\begin{figure}[t]
\centering
\includegraphics[width=0.9\textwidth]{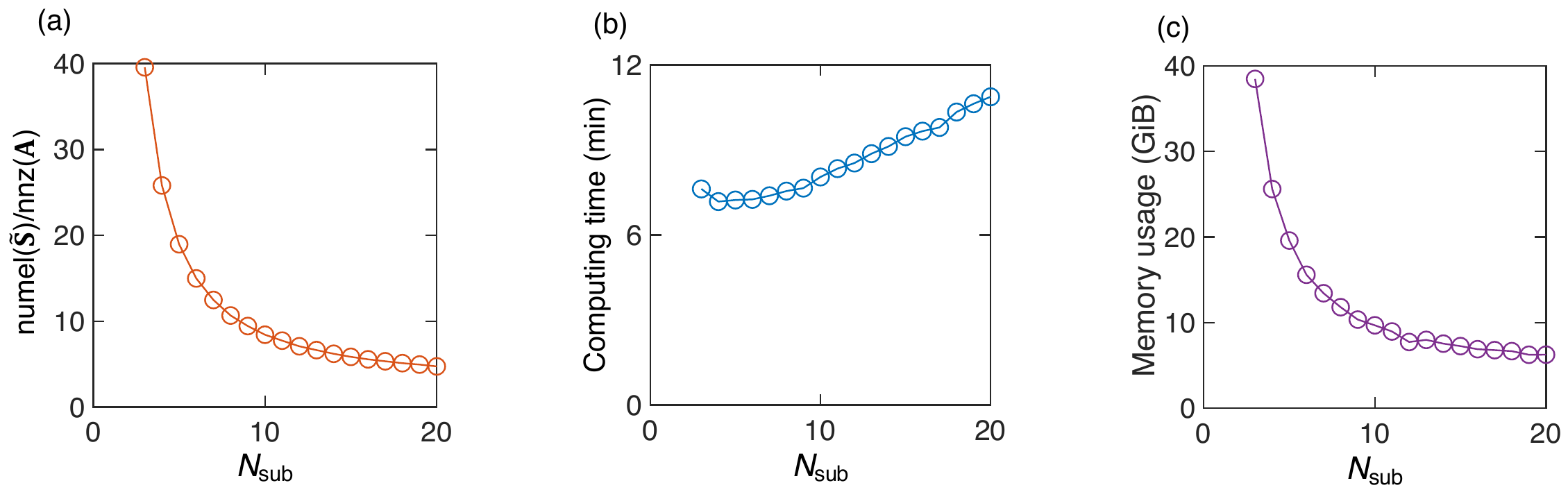}
\caption{(a) The size of matrix $\bf \Tilde{S}$ compared to the number of nonzero elements in matrix $\bf A$ when one objective-plus-gradient APF computation is divided into $N_{\rm sub}$ sub-ones for a 1-mm-wide metasurface.
The (b) total computing time and (c) memory usage as a function of $N_{\rm sub}$.}
\label{fig:benchmark}
\end{figure}
For the APF method,
%The sparse linear solver package MUMPS~\cite{amestoy2001fully} is used to efficiently perform the partial factorization in APF, which benefits from the symmetry of the problem.
we build matrices $\bf B$ and $\bf C$ using channels from both two sides of the metasurface and retrieve responses of desirable channels after the partial factorization to take full advantage of the symmetry of the augmented matrix ${\bf K}=[{\bf A},{\bf B};{\bf C},{\bf D}]$ and reduce the computing time and memory usage.
Note that here nnz($\bf B$) $\approx 2.3\times 10^8$ exceeds nnz($\bf A$) $\approx 1.6\times 10^7$,
thus we compress $\bf B$ and $\bf C$ to reduce the number of nonzero elements such that nnz($\bf B$) < nnz($\bf A$) and the computational cost does not grow with $M_\mathrm{in}$~\cite{lin2022_APF}.
Specifically, we pad 480 extra channels to the input/output profiles, weight them by the Hann window function, Fourier transform them to the spatial domain,  and then truncate with a 3$\lambda$ window.
After performing the partial factorization, the output is decompressed by undoing the above process, with a negligible error of $\sim10^{-4}$ from compression.
%When the gradient information is required, we build and compute ${\bf \Tilde{S}}={\bf \tilde C}{\bf A}^{-1}{\bf \tilde B}=[{\bf C};{\bf U}^{\rm T}]{\bf A}^{-1}[{\bf B},{\bf U}]$, which ensures the augmented matrix ${\bf \tilde K}=[{\bf A},{\bf \tilde B};{\bf \tilde C},{\bf D}]$ to be square and symmetric.

For the conventional method, we factorize matrix $\mathbf{A}=\mathbf{LU}$, solve $\mathbf{A}^{-1}B_m$ for each input $B_m$ and project it with $\mathbf{C}$ if only $f$ is needed.
To compute the gradient using adjoint method, we perform the forward simulation (solve $\mathbf{A}^{-1}B_m$), and the successive adjoint simulation with source $C_m^\mathrm{adj}=\sum_{n=1}^{M_\mathrm{out}}(\partial f/\partial S_{nm})C_n$ (solve $C_m^\mathrm{adj}\mathbf{A}^{-1}$) for each input $B_m$.
The gradient is obtained after a loop over all $M_\mathrm{in}$ inputs:
\begin{equation}
    \frac{df}{dp_k} = \frac{\partial f}{\partial p_k} - \sum_{m=1}^{M_\mathrm{in}} 2\mathrm{Re} \left( C_m^\mathrm{adj}\mathbf{A}^{-1}\frac{\partial \mathbf{A}}{\partial p_k} \mathbf{A}^{-1}B_m \right).
    \label{eq:grad_adj}
\end{equation}

The gradient computation code with these two methods implemented is open-source on Ref.~\cite{APF_inv_github}.
All computations are performed using a single core on an Intel Xeon 2640v4 node.
Each case is computed 10 times, with the average computing time and the maximal memory usage after subtracting the 0.9 GiB memory used by MATLAB R2022a recorded. As shown in Tab.~\ref{tab:benchmark_1200},
%The total computing time is broken down into time used in building corresponding matrices ({\it i.e.}, $\bf A$, $\bf B$, $\bf C$, and augmented matrix $\bf K$ for APF only), analysis, factorization, solving (conventional method only), and post-processing procedure needed to get the accurate scattering matrix (APF only).
factorization and solving are the most time-consuming stages for the APF and conventional method, respectively.
Extra timing needed to obtain $f$ and $\bm{\nabla}_\mathbf{p} f$ from the full-wave simulations is counted in post-processing.
APF outperforms the conventional method by $\sim$2000 fold in speed and $\sim$2.5 fold in memory for the $f$-only case, with the memory usage difference coming from the compression of $\mathbf{B}$ and $\mathbf{C}$ in APF.
When gradient is involved, 
APF can still outperform the conventional adjoint method in both speed and memory by separating a single computation into $N_\mathrm{sub}$ ones, as shown in Fig.~\ref{fig:benchmark} with $N_\mathrm{sub} \ge 3$ limited by the memory we have access to.
One can choose $N_\mathrm{sub}$ to match their major concerning.
As presented in Tab.~\ref{tab:benchmark_1200}, APF with $N_\mathrm{sub}=15$ achieves a $\sim$150 times speed-up and a $\sim$30\% memory usage reduction compared to the conventional adjoint method.

\section{Meta-atoms of the metasurface}
\label{sec:meta_atom}

\begin{figure}[htbp]
\centering
\includegraphics[width=0.8\textwidth]{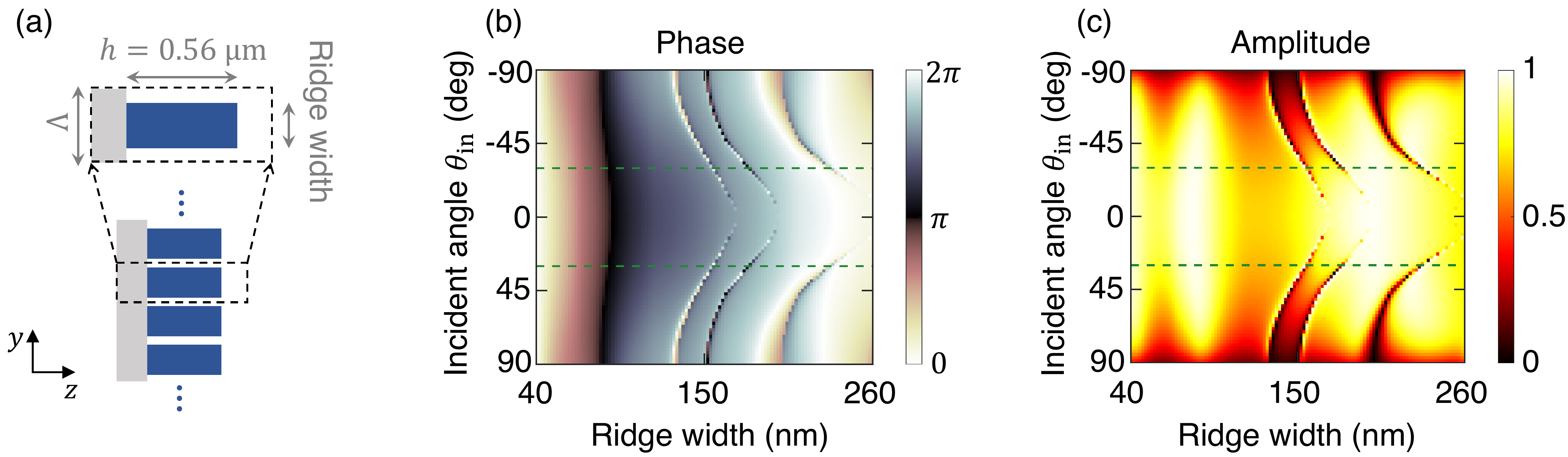}
\caption{(a) Schematic structure of a periodic array of unit cells.
    (b) Phase and (c) amplitude maps of the zeroth-order transmission coeﬃcient for diﬀerent ridge widths (from 40 nm to 260 nm) and incident angles.    
    The green dashed lines mark the angular range of interest $\theta_{\rm in} \in [-30^{\circ},30^{\circ}]$.}
\label{fig:meta_atom_fig}
\end{figure}

Figure~\ref{fig:meta_atom_fig} shows that ridges with height $h=0.56$ \textmu m can provide a $2\pi$ range of phase shift with high transmission over the 60$^{\circ}$ angular range of interest by changing their widths.
We obtain the response of each unit cell with full-wave simulations using APF method implemented in MESTI~\cite{MESTI},
%Each unit cell is discretized with grid size $\Delta x=\lambda/40$, 
adopting Bloch periodic boundary in $y$ and 20 pixels of perfectly matched layers (PMLs) in $z$ to mimic a periodic array of meta-atoms with a fixed periodicity of $\Lambda=300$ nm.
Then the phase and amplitude of the zeroth-order transmission coefficient are mapped out for different ridge widths and incident angles.

\section{Gradient validation}
\label{sec:grad_valid}

\begin{figure}[htbp]
\centering
\includegraphics[width=0.55\textwidth]{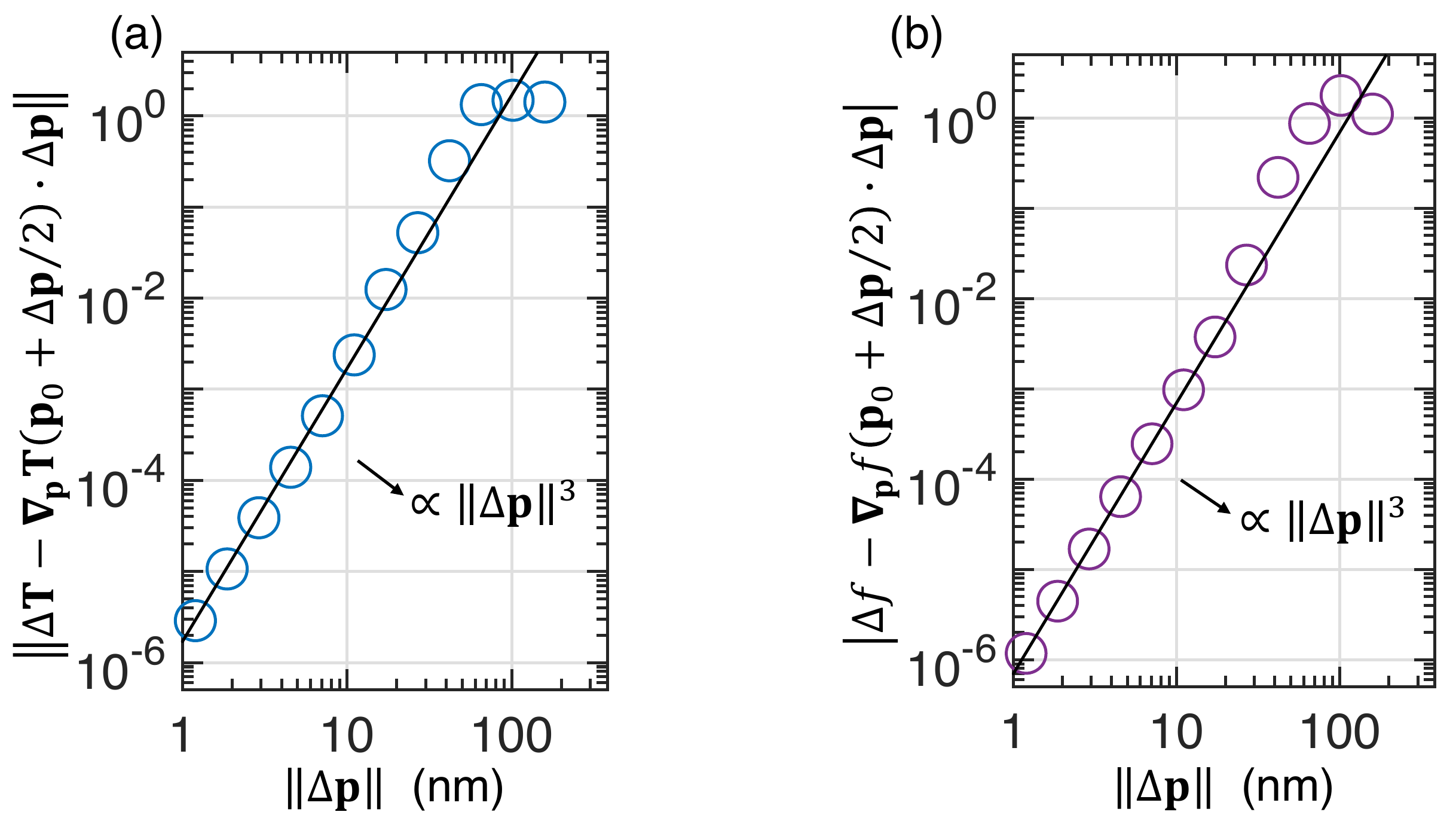}
\caption{Log-log plots of the difference (a) $||\bm{\Delta}\mathbf{T}-\bm{\nabla}_{\bf p} \mathbf{T}({\bf p}_0+\Delta {\bf p}/2)\cdot \Delta {\bf p}||$ and (b) $|\Delta f - \bm{\nabla}_{\bf p} f({\bf p}_0+\Delta {\bf p}/2) \cdot \Delta {\bf p}|$ as a function of the increment of optimization variables $||\Delta {\bf p}||$, where $\bm{\Delta}\mathbf{T} = \mathbf{T}({\bf p}_0+\Delta {\bf p})-\mathbf{T}({\bf p}_0)$ and $\Delta f =f({\bf p}_0+\Delta {\bf p}) - f({\bf p}_0)$. The solid black lines have a slope of 3, indicating that the difference scales as $\mathcal{O}(||\Delta {\bf p}||^3)$ for small $||\Delta {\bf p}||$.}
\label{fig:grad_valid}
\end{figure}

To validate the gradient derivation and implementation, Figure~\ref{fig:grad_valid} compares the gradient computed with APF to finite-difference evaluation for a metasurface composed of $N=80$ ridges described in Fig.~2b,c of the main text, with the FoM $f$ defined in Eq.~(5).
The transmission matrix $\mathbf{T}$ and FoM $f$ are evaluated at ${\bf p}_0$ and ${\bf p}_0+\Delta {\bf p}$, where $\Delta {\bf p}$ is a random increment of design parameters.
%generated by multiplying different prefactors $q$ to a list of random numbers between $[-1,1]$ with an $l^2$-norm of 1, thus$||\Delta {\bf p}||=q$.
Taylor expansion shows that the $l^2$-norm $||\bm{\Delta}\mathbf{T}-\bm{\nabla}_{\bf p} \mathbf{T}({\bf p}_0+\Delta {\bf p}/2)\cdot \Delta {\bf p}||$ with $\bm{\Delta}\mathbf{T} = \mathbf{T}({\bf p}_0+\Delta {\bf p})-\mathbf{T}({\bf p}_0)$ should scale as $\mathcal{O}(||\Delta {\bf p}||^3)$ when $||\Delta {\bf p}||$ is small, if the gradient $\bm{\nabla}_{\bf p} \mathbf{T}({\bf p}_0+\Delta {\bf p}/2)$ is accurately computed; so does the difference $|\Delta f - \bm{\nabla}_{\bf p} f({\bf p}_0+\Delta {\bf p}/2) \cdot \Delta {\bf p}|$ with $\Delta f =f({\bf p}_0+\Delta {\bf p}) - f({\bf p}_0)$, which is observed in Fig.~\ref{fig:grad_valid}.
%As shown in Fig.~\ref{fig:Fig2}, the $\mathcal{O}(||\Delta P||^3)$ scaling is observed when $||\Delta P||\lesssim$ 10 nm, indicating that the gradients are correctly computed.

\section{Compare different optimization algorithms}
\label{sec:opt_methods}

\begin{figure}[htbp]
\centering
\includegraphics[width=0.47\textwidth]{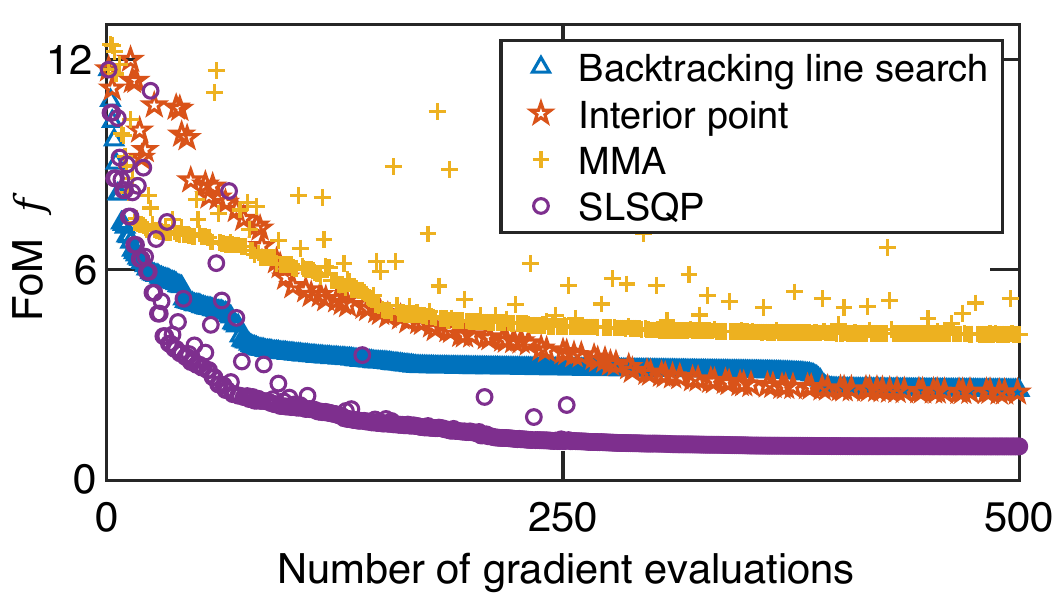}
\caption{The FoM evolves as the optimization goes on using different algorithms.}
\label{fig:FoM_opt_methods}
\end{figure}

Different well-developed gradient-based algorithms can be used to optimize the structure parameters $\{ p_k\} = \{ y_1,\cdots,y_N\}$ and minimize the FoM.
Figure~\ref{fig:FoM_opt_methods} compares how the FoM evolves as the optimization goes on using different algorithms, starting from the same initial guess.
%During the optimizations, we constrain neighboring $y_n$ to be separated by at least 40 nm to ensure fabrication feasibility.

An intuitive choice is to update $\mathbf{p}$ along the opposite direction of $\bm{\nabla}_{\bf p} f$ by an adjustable step as ${\bf p}_{n+1} = {\bf p}_n - \gamma_n \bm{\nabla}_{\bf p} f({\bf p}_n)$, where $\gamma_n$ is the learning rate at the $n$-th iteration; $n$ represents the iteration number.
The learning rate $\gamma_n$ is chosen using the ``backtracking line search'' method~\cite{nocedal2006line,truong2021backtracking} so as not to overshoot (when $\gamma_n$ is too large) or to slow down convergence (when $\gamma_n$ is too small).
Here in Fig.~\ref{fig:FoM_opt_methods}, we start from a relatively large estimate of the learning rate $\gamma_{\rm max}=2\times 10^{-4}$ \textmu m$^2$ for movement along the $-\bm{\nabla}_{\bf p} f$ direction, and iteratively shrink the learning rate by a factor of $\tau=1/2$ until the Armijo condition~\cite{armijo1966minimization} is satisfied:
\begin{equation}
    f[{\bf p}_n-\gamma_n \bm{\nabla}_{\bf p} f({\bf p}_n)] \leq f({\bf p}_n) - c_1\gamma_n [\bm{\nabla}_{\bf p} f({\bf p}_n)]^{\rm T} \bm{\nabla}_{\bf p} f({\bf p}_n),
\label{eq:Armijo}
\end{equation}
where $c_1=10^{-2}$.

We also try the sequential least-squares quadratic programming (SLSQP) algorithm~\cite{kraft1994algorithm} and method of moving asymptotes (MMA)~\cite{svanberg2002class} implemented in the open-source package NLopt~\cite{NLopt}, and interior point algorithm~\cite{byrd1999interior,byrd2000trust,waltz2006interior} implemented in the fmincon function in Matlab; all of them support inequality constraints.
As shown in Fig.~\ref{fig:FoM_opt_methods}, SLSQP converges faster and to a lower level among all these algorithms, and thus is adopted for all optimizations in this work.

Note that although the gradient $-\bm{\nabla}_{\bf p}f$ points out the update direction, one still needs intermediate steps to decide the proper update size.
A large update may cause overshooting; this is what happens at some iterations for interior point algorithm, MMA and SLSQP in Fig.~\ref{fig:FoM_opt_methods}.
In the backtracking line search algorithm, we skip intermediate steps and only record evaluations that reduce the FoM, so the curve keeps going down.

\begin{comment}
\section{Initial FoM and final FoM}
\label{sec:initial_final_FoM}

\begin{figure}[htbp]
\centering
\includegraphics[width=0.4\textwidth]{initial_final_FoM_fig.pdf}
\caption{The corresponding final FoM as a function of 1000 initial guesses for symmetric metasurfaces.}
\label{fig:initial_final_FoM_fig}
\end{figure}

The performance of gradient-based algorithms depends strongly on the initial guess of the solution,
and an appropriately chosen initial design may lead to a globally good result.
Figure~\ref{fig:initial_final_FoM_fig} shows no relation between the initial FoM and final FoM, which implies that starting from a small FoM may not end at a small FoM.
We still needs to generate plenty of initial designs, perform optimizations, and pick the optimal one from them.
\end{comment}

\section{Evolution of the optimization (animation)}
\label{sec:animation_trans_matrix}

\begin{figure}[htbp]
\centering
\includegraphics[width=0.5\textwidth]{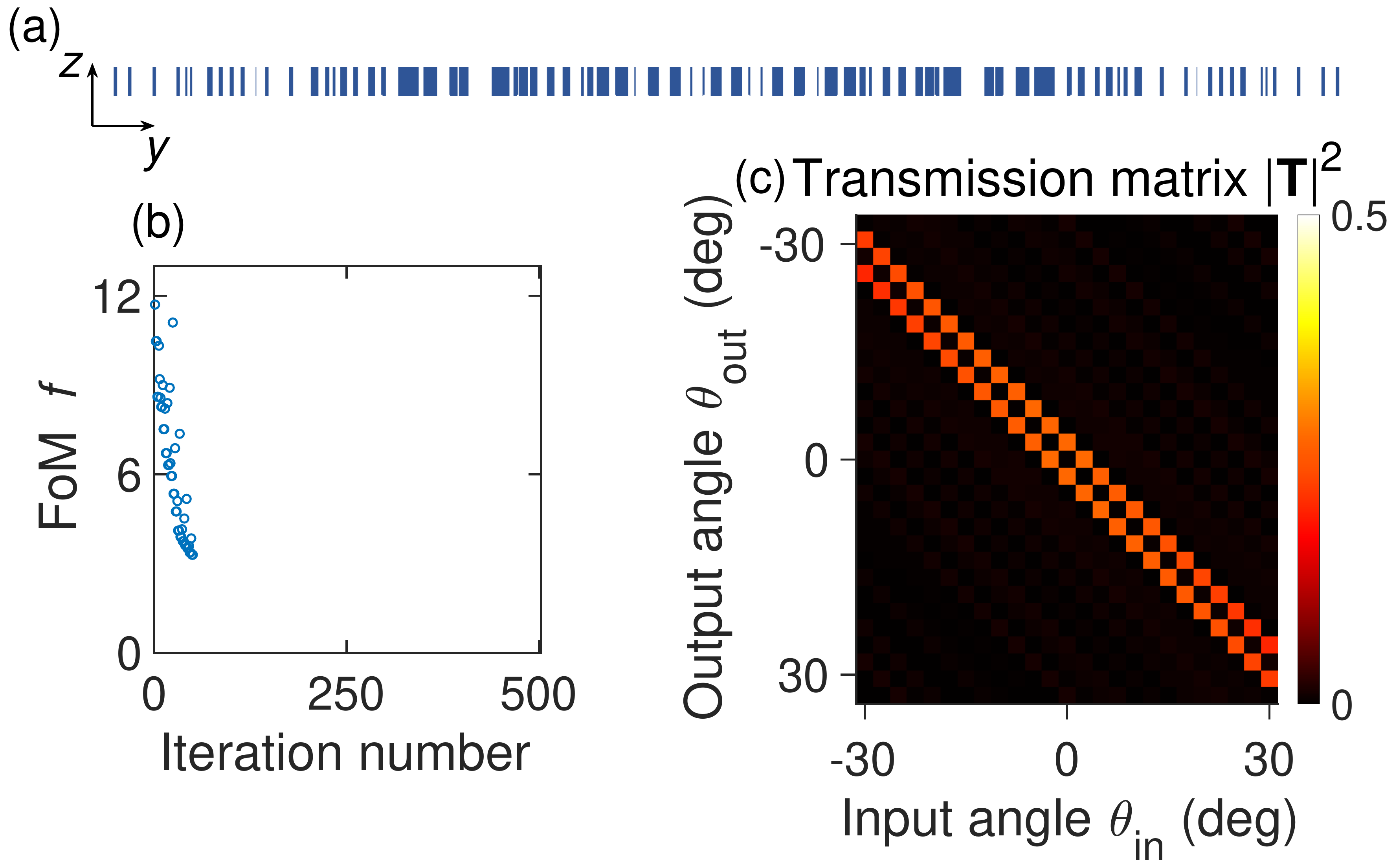}
\caption{One frame of Supplementary Video 1, which shows how the metasurface, its transmission matrix and the FoM evolve as the optimization goes on.
(a) Metasurface and (c) transmission matrix $|\mathbf{T}|^2$ at the 50-th iteration.
(b) Evolution of the FoM $f$ during the first 50 iterations.}
\label{fig:animation_frame}
\end{figure}

Supplementary Video~1 shows how the metasurface, its transmission matrix $|{\bf T}|^2$ and the FoM evolve as the optimization goes on.
Figure~\ref{fig:animation_frame} provides the animation caption and shows one frame of the animation.

The edge positions ${\bf p} = \{ p_k\} = \{ y_1,\cdots,y_{80}\}$ with $\{ y_{81},\cdots,y_{160}\}=-\{ y_{80},\cdots,y_1\}$ for the optimized 80-ridge metasurface are listed in Tab.~\ref{tab:edge_opt}.
\begin{table}[htbp]
\newcommand{\tabincell}[2]{\begin{tabular}{@{}#1@{}}#2\end{tabular}}
\centering
\caption{\bf Edge positions of the optimized metasurface (in micron)}
\begin{tabular}{lc}
\toprule
    $\{ y_1, \cdots, y_8 \}$ &  -11.863, -11.778, -11.497, -11.413, -11.131, -11.047, -10.791, -10.708\\
\cmidrule(rr){1-1} \cmidrule(rr){2-2}
     $\{ y_9, \cdots, y_{16} \}$ & -10.503, -10.463, -10.423, -10.357, -10.111, -10.026, -9.769, -9.695\\
\cmidrule(rr){1-1} \cmidrule(rr){2-2}
    $\{ y_{17}, \cdots, y_{24} \}$ & -9.581, -9.512, -9.300, -9.221, -9.130, -9.086, -8.927, -8.843\\
\cmidrule(rr){1-1} \cmidrule(rr){2-2}
    $\{ y_{25}, \cdots, y_{32} \}$ & -8.557, -8.472, -8.183, -8.097, -7.806, -7.730, -7.674, -7.632\\
\cmidrule(rr){1-1} \cmidrule(rr){2-2}
    $\{ y_{33}, \cdots, y_{40} \}$ & -7.489, -7.361, -7.203, -7.111, -6.952, -6.814, -6.648, -6.560\\
\cmidrule(rr){1-1} \cmidrule(rr){2-2}
    $\{ y_{41}, \cdots, y_{48} \}$ & -6.317, -5.891, -5.753, -5.530, -5.354, -5.151, -5.111, -4.963\\
\cmidrule(rr){1-1} \cmidrule(rr){2-2}
    $\{ y_{49}, \cdots, y_{56} \}$ & -4.577, -4.267, -4.103, -4.047, -4.007, -3.796, -3.741, -3.633\\
\cmidrule(rr){1-1} \cmidrule(rr){2-2}
    $\{ y_{57}, \cdots, y_{64} \}$ & -3.593, -3.452, -3.170, -3.007, -2.956, -2.916, -2.701, -2.575\\
\cmidrule(rr){1-1} \cmidrule(rr){2-2}
    $\{ y_{65}, \cdots, y_{72} \}$ & -2.529, -2.306, -2.137, -1.905, -1.678, -1.635, -1.432, -1.249\\
\cmidrule(rr){1-1} \cmidrule(rr){2-2}
    $\{ y_{73}, \cdots, y_{80} \}$ & -1.179, -0.964, -0.695, -0.655, -0.443, -0.400, -0.322, -0.094\\
\bottomrule
\end{tabular}
  \label{tab:edge_opt}
\end{table}

\section{Optimization without mirror symmetry}
\label{sec:general}

\begin{figure}[htbp]
\centering
\includegraphics[width=0.7\textwidth]{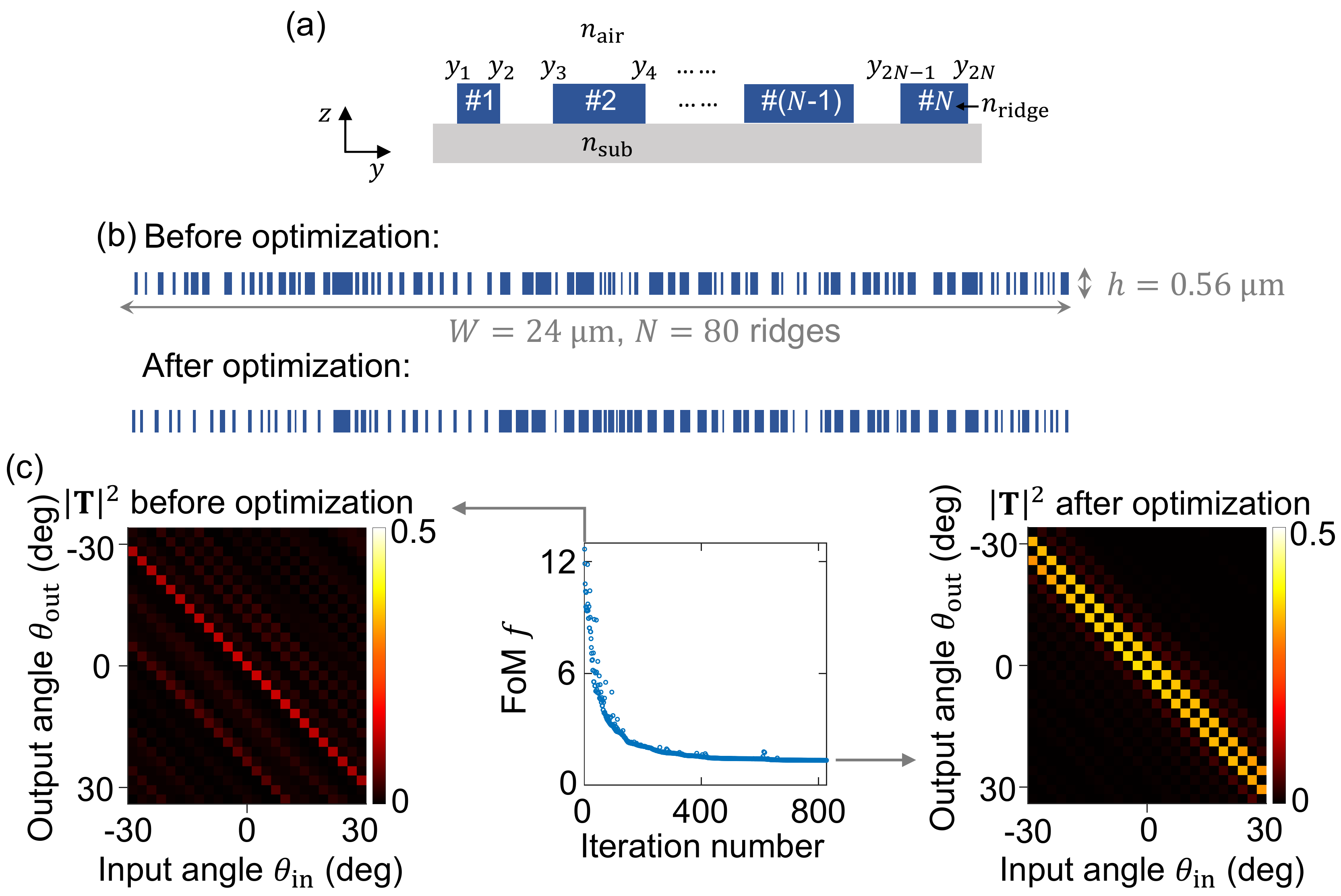}
\caption{Schematic of (a) a general metasurface without mirror symmetry.
    (b) Metasurfaces before and after optimization.
    (c) Evolution of the FoM $f$ during optimization and the squared amplitude of transmission matrices before and after optimization.}
\label{fig:general_best}
\end{figure}

\begin{figure}[htbp]
\centering
\includegraphics[width=0.6\textwidth]{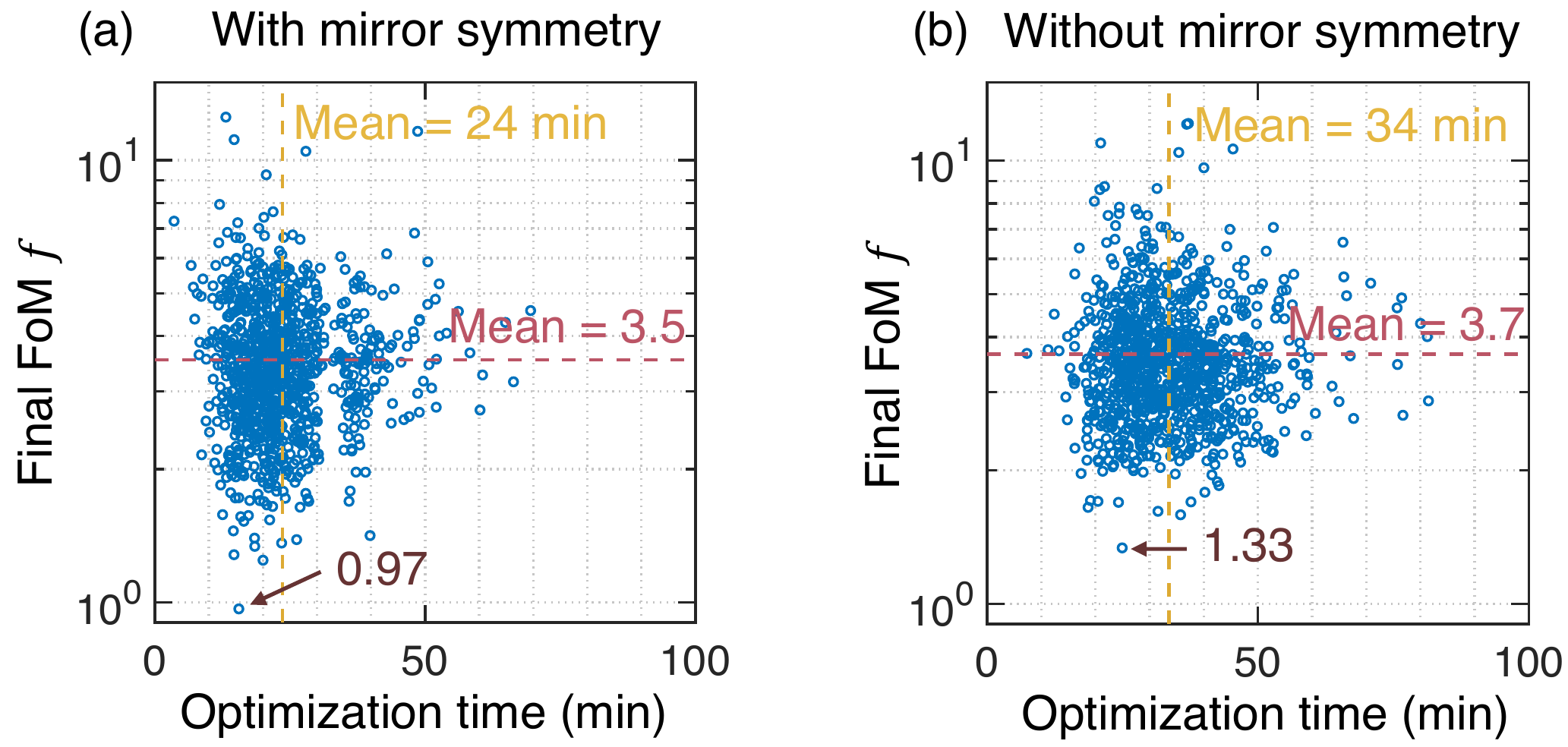}
\caption{The final FoM and optimization time starting from 1000 different initial guesses with dashed lines representing the average FoM (red) and optimization time (yellow), for metasurfaces (a) with mirror symmetry and (b) without mirror symmetry. The minimal FoM achieved over these 1000 optimizations is marked by an arrow.}
\label{fig:FoM_time}
\end{figure}

Figure~2 of the main text shows the optimized broad-angle metasurface beam splitter with mirror symmetry.
%based on symmetric metasurfaces with edge positions of ridges satisfying $\{ y_{N+1},\cdots,y_{2N}\}=-\{ y_N,\cdots,y_1\}$.
Here, we optimize metasurfaces without mirror symmetry (update all edge positions $\{ y_1,\cdots,y_{2N}\}$ independently),
and provide structures and transmission matrices before and after optimization in Fig.~\ref{fig:general_best}.
The optimal metasurface over 1000 initial guesses has $|T_{nm}|^2$ averaged to 0.3 at the $\pm 1$ diffraction orders over the $60^{\circ}$ angular range.
The total 827 iterations cost 25 minutes, yielding a final FoM of 1.33.
As shown in Fig.~\ref{fig:FoM_time},
metasurfaces without mirror symmetry typically yield less optimal results compared to those with mirror symmetry.

%\begin{acknowledgement}

%Please use ``The authors thank \ldots'' rather than ``The
%authors would like to thank \ldots''.

%The author thanks Mats Dahlgren for version one of \textsf{achemso},
%and Donald Arseneau for the code taken from \textsf{cite} to move
%citations after punctuation. Many users have provided feedback on the
%class, which is reflected in all of the different demonstrations
%shown in this document.

%\end{acknowledgement}

%\begin{suppinfo}

%This will usually read something like: ``Experimental procedures and
%characterization data for all new compounds. The class will
%automatically add a sentence pointing to the information on-line:

%\end{suppinfo}

\bibliography{bib_supp_APF_inverse_design}